\documentclass[aps,pre,reprint,amsmath,amssymb,superscriptaddress,showpacs,longbibliography]{revtex4-1}

\usepackage{bm}
\usepackage{graphicx}
\usepackage{xcolor}
\usepackage{url}
\usepackage[normalem]{ulem}
\usepackage{times}
\usepackage{bbm}
\usepackage{mathrsfs} 
\usepackage{comment}
\usepackage{dsfont}
\usepackage{pdfpages}
\usepackage{bbm} 
\def\1{\ensuremath{{\mathbbm{1}}}}
\usepackage[colorlinks=true,breaklinks=true,allcolors=blue]{hyperref}
\usepackage{url}

\usepackage[normalem]{ulem}

\definecolor{michael}{rgb}{0.2,.6,.7}

\definecolor{hannes}{rgb}{0.4,0.4,0}

\definecolor{daniel}{rgb}{0, 0, 0}
\definecolor{jan}{RGB}{240, 128, 128}

\newcommand{\ex}[1]{\ensuremath{\left\langle #1 \right\rangle}}
\newcommand{\mvec}[1]{\boldsymbol #1}
\newcommand{\ev}[1]{\langle #1 \rangle}
\newcommand{\Ldag}{\mathcal{L}^\dagger}

\newcommand\numberthis{\addtocounter{equation}{1}\tag{\theequation}} 
\DeclareMathOperator{\Tr}{{Tr}}
\DeclareMathOperator{\myRe}{{Re}}
\DeclareMathOperator{\myIm}{{Im}}
\DeclareMathOperator{\cn}{{cn}}
\def\Oo{\ensuremath{{\cal O}}} 
\def\Uu{\ensuremath{{\cal U}}} 
\def\Dd{\ensuremath{{\cal D}}} 
\def\Ll{\ensuremath{{\cal L}}} 
\def\Hh{\ensuremath{{\cal H}}} 

\makeatletter
\let\cat@comma@active\@empty
\makeatother

\makeatletter
\AtBeginDocument{\let\LS@rot\@undefined}
\makeatother

\begin{document}

\title{Thermalization of a Lipkin-Meshkov-Glick model coupled to a bosonic bath}

\author{Jan C.\ Louw}
\affiliation{Institute of Theoretical Physics,  Department of Physics, University of Stellenbosch, Stellenbosch 7600, South Africa}

\author{Johannes N.\ Kriel}
\affiliation{Institute of Theoretical Physics,  Department of Physics, University of Stellenbosch, Stellenbosch 7600, South Africa}

\author{Michael Kastner}
\email{kastner@sun.ac.za}
\affiliation{National Institute for Theoretical Physics (NITheP), Stellenbosch 7600, South Africa}
\affiliation{Institute of Theoretical Physics,  Department of Physics, University of Stellenbosch, Stellenbosch 7600, South Africa}

\date{\today}

\begin{abstract}
We derive a Lindblad master equation that approximates the dynamics of a Lipkin-Meshkov-Glick (LMG) model weakly coupled to a bosonic bath. By studying the time evolution of operators under the adjoint master equation we prove that, for large system sizes, these operators attain their thermal equilibrium expectation values in the long-time limit, and we calculate the rate at which these values are approached. Integrability of the LMG model prevents thermalization in the absence of a bath, and our work provides an explicit proof that the bath indeed restores thermalization. Imposing thermalization on this otherwise non-thermalizing model outlines an avenue towards probing the unconventional thermodynamic properties predicted to occur in ultracold-atom-based realizations of the LMG model.
\end{abstract}

\maketitle

\section{Introduction}

General experience suggests that, after a sufficiently long time, an isolated system approaches thermal equilibrium, characterized by its initial energy but not by other details of the initial state. While occurrence of thermalization is widespread, it is also well established that certain classes of systems defy thermalization and preserve a more detailed memory of their initial conditions. Prominent examples are integrable systems, possessing an extensive number of (quasi)local conserved quantities \cite{Rigol_etal07,VidmarRigol16}, or many-body localized systems, where disorder effectively generates conserved quantities \cite{NandkishoreHuse15,GogolinEisert16}. When a system is not isolated but coupled to an infinite bath, thermalization to a Gibbs state characterized by the temperature of the bath seems even more plausible, but a general proof of such behavior is lacking.

In the absence of a general proof we study a specific model that, due to integrability, does not thermalize when isolated from its environment, and we investigate whether a bath can indeed facilitate thermalization. The model we consider was introduced by Lipkin, Meshkov, and Glick (LMG) \cite{LipkinMeshkovGlick65} in the context of nuclear physics, and has subsequently found applications also in other branches of physics. The model can be written either as a bosonic Hamiltonian or, via a Schwinger mapping, in terms of pseudo-spin operators \cite{LipkinMeshkovGlick65}. Our interest in this model is motivated by experiments with ultracold atoms for which it was shown that, under suitable circumstances, a two-component Bose--Einstein condensate can be described by an LMG model \cite{Zibold_etal10,Strobel_etal14}. In the experimental realizations, the bosonic LMG Hamiltonian is restricted to a Fock space sector of fixed boson number $N$, corresponding to a fixed spin quantum number $S=N/2$ in the pseudo-spin representation \eqref{e:Hspin} of the model. Because of this restriction the Hilbert space dimension of the system does not grow exponentially, but only linearly with the system size $N$ or, equivalently, the spin quantum number $S$. A number of unconventional thermodynamic properties have been shown to arise from this subexponential (linear) growth of the Hilbert space, including distribution functions that do not concentrate in the large-$S$ limit \cite{WebsterKastner18}. 

In the abovementioned experimental realizations, the LMG model is well isolated from its environment and evolves under unitary dynamics. Since the model is solvable via Bethe Ansatz \cite{Ortiz_etal05}, its equilibrium is in general not described by a thermal state. An experimental realization of a bath that couples to the LMG system in a controlled way, even though possibly a difficult task, may be feasible in one way or the other \cite{WebsterKastner18}. In this paper we theoretically model and analyze this situation by considering an LMG model coupled to an infinitely large bath of independent bosons, characterized by some spectral density. This bath, and also the form of the system--bath coupling, is not necessarily a faithful modeling of a specific future experimental realization. However, on the theoretical side, certain qualitative features of the dynamics are believed to be rather insensitive to the details of the bath. This creates the expectation that what we learn in the context of the specific bath introduced in Sec.~\ref{s:model} might be valid for a broader class of environments, hopefully including those that are relevant for potential future experimental realizations of an LMG system coupled to a bath.

In this paper, we use an LMG model (the system) weakly coupled to a bosonic bath (the bath) as a starting point for deriving, under fairly standard assumptions \cite{BreuerPetruccione}, a Markovian master equation of Lindblad form \cite{Lindblad76} that describes the time evolution of the system's density operator, while the bath degrees of freedom have been traced out. The main body of work that we want to report in this paper then consists of analyzing the properties of this master equation, and in particular the questions of whether a thermal state is approached in the long-time limit and, if so, at what rate. We address these questions by using the adjoint master equation, which governs the Lindblad time evolution of operators in the Heisenberg picture. By exploiting the presence of widely separated time scales, the main technical achievement of this paper is an asymptotic solution of those time-evolution equations for a relevant set of observables in the large-system limit.

Our main finding is that the presence of a bosonic bath indeed induces thermalization of the integrable LMG model for the expectation values of the system Hamiltonian density as well as of all three spin component densities $S_x/S$, $S_y/S$, and $S_z/S$, independently of the initial state of the system. Moreover, our results confirm, from an open-systems perspective, an unconventional aspect of the equilibrium ensemble calculation of Ref.~\cite{WebsterKastner18}, namely the temperature independence of thermal expectation values in the large-$S$ limit. In particular, when taking the infinite-system limit at any fixed bath temperature $T$, the expectation values of the Hamiltonian and spin component densities equilibrate to their ground state values. This unconventional thermal behavior is a consequence of the subexponential growth of the Hilbert space dimension of the LMG model with $S$, which essentially prevents entropic effects from influencing the thermal properties; see \cite{WebsterKastner18} for details. In addition to establishing thermalization, we calculate in Sec.~\ref{s:approach} the dissipation rate at which thermal equilibrium is reached, which is found to depend on the (quantum) phase to which the equilibrium state belongs. To induce thermalization to an equilibrium state which yields expectation values different from that of the ground state, it is necessary to take the infinite-system limit while keeping the rescaled temperature $\widetilde{T}\equiv T/S$ fixed. We consider this case in Sec.~\ref{ss:ExT} and provide evidence of thermalization on an observable-independent level by showing that a canonical Gibbs state is the unique stationary state of the Lindblad master equation.

\section{Model}
\label{s:model}

The LMG model was proposed by Lipkin, Meshkov, and Glick in 1965 as a Hamiltonian that describes interacting bosons. In that same paper it was also recognized that, by means of a Schwinger mapping, the bosonic Hamiltonian can be expressed in terms of spin operators. Here we consider a slightly less general version of that model, which in spin language can be written as
\begin{equation}\label{e:Hspin}
\mathcal{H}_S=-\frac{\Lambda}{2S}S_x^2-S_z,
\end{equation}
where $S_x$ and $S_z$ are the $x$- and $z$-components of a spin-$S$ vector operator, and $\Lambda$ is a coupling constant. The number of particles $N$ in the bosonic formulation of the model is related to the spin quantum number $S$ in the spin formulation via $N=2S$. In later sections of this paper we will consider the LMG model in the limit $S\to\infty$ of infinite boson number or, equivalently, the semi-classical limit of infinite spin quantum number $S$. This model exhibits a quantum phase transition at $\Lambda=1$ from the symmetric to the symmetry-broken phase. In the former the model has a non-degenerate ground state with a zero $S_x$ expectation value. In the limit $N\to\infty$ the ground state in the symmetry-broken phase will be two-fold degenerate with non-zero $S_x$ expectation values \cite{DusuelVidal05}.

We consider the prototypical environment of a bosonic bath Hamiltonian
\begin{equation}
\mathcal{H}_B=\sum_q \omega_q a_q^\dagger a_q, 
\end{equation}
where $a_q$ and $a_q^\dagger$ are bosonic ladder operators and $\omega_q$ are the mode frequencies. The total system is described by the Hamiltonian
\begin{equation}
\mathcal{H}=\mathcal{H}_S+\mathcal{H}_B+\mathcal{V},
\end{equation}
where
\begin{equation}
\mathcal{V}=\sqrt{\frac{\gamma}{S}}\,S_x\otimes\mathcal{B} \label{eq:coup}
\end{equation}
with
\begin{equation}
\mathcal{B}=\sum_q \gamma_q\left(a_q^\dagger+a_q\right)
\end{equation}
couples the system to the bath. The coupling strength between system and bath is determined by $\gamma$, which we will later on consider as small. For specific parameter values, $\mathcal{H}$ reduces to models previously studied in the literature. For $S=1/2$ one obtains the spin--boson model \cite{Leggett_etal87}, while for $\Lambda=0$ one obtains the multi-modal Dicke Hamiltonian \cite{HeppLieb73}.

The form of the coupling operator \eqref{eq:coup} is a convenient choice for the calculations. We expect, but have not shown, that the main conclusions of this paper remain valid for other coupling directions, like an $S_y$ operator in \eqref{eq:coup}. An exception is a coupling operator \eqref{eq:coup} with $S_z$ in the place of $S_x$. In this case the total Hamiltonian conserves the parity symmetry of the spin, implying that the bath can destroy only some of the integrals of motion of the system, but not all of them, and hence thermalization cannot be expected to occur.

\section{Lindblad master equation}
\label{s:MasterEquation}
To arrive at a more tractable problem, we use standard approximations, detailed in the Appendices \ref{A:BMM} and \ref{ssec:pos}, to derive a Born-Markov master equation of Lindblad form that approximately describes the time evolution of the density operator $\rho_\text{S}$ of the LMG model weakly coupled to a bath,
\begin{equation}\label{e:MasterGeneral}
\partial_t \rho_\text{S} = \mathcal{L}\rho_\text{S}
\end{equation}
with
\begin{align}\label{e:Lindbladian}
\mathcal{L}\bullet=&i[\bullet,\mathcal{H}_S]-i\myIm(\kappa_{yx})\,[S_x,\{S_y,\bullet\}]-\frac{\kappa_{xx}}{2}[S_x,[S_x,\bullet]]\nonumber\\
&-\frac{|\kappa_{xy}|^2}{\kappa_{xx}}[S_y,[S_y,\bullet]]+2\myRe(\kappa_{xy})[S_y,[S_x,\bullet]].
\end{align}
The entries of the $\kappa$ matrix are given in Appendix \ref{ssec:pos}. As part of the approximations when deriving the Lindblad master equation, the bath is at all times assumed to be in the Gibbs state $\rho_{B}\propto\exp\left(-\mathcal{H}_B/T\right)$ with temperature $T$ as well as uncorrelated with the system. Equation \eqref{e:MasterGeneral} describes the time evolution of the state (or density operator) of an open quantum system, analogous to the Schr\"odinger equation for a closed system. For closed systems it is well-known that, instead of time-evolving the state, an equivalent description of the quantum dynamics is possible in the Heisenberg picture, where operators evolve and states are constant in time. The same change of viewpoint is possible for open quantum systems, where the adjoint master equation 
\begin{equation}\label{e:AdjointMasterEquation}
\partial_t A(t)=\Ldag A(t)
\end{equation}
describes the time evolution of an observable $A$ under the Lindblad dynamics of an open quantum system. Here $\Ldag$ denotes the adjoint Lindbladian \cite{BreuerPetruccione}
\begin{align}\label{e:AdjointLindbladian}
\Ldag \bullet=&i[\mathcal{H}_S,\bullet]+i\myIm(\kappa_{yx})\,\{S_y,[S_x,\bullet]\}-\frac{\kappa_{xx}}{2}[S_x,[S_x,\bullet]]\nonumber\\
&-\frac{|\kappa_{xy}|^2}{\kappa_{xx}}[S_y,[S_y,\bullet]]+2\myRe(\kappa_{xy})[S_x,[S_y,\bullet]],
\end{align}
which satisfies $\Tr[A(\mathcal{L}\rho_S)]=\Tr[(\Ldag A)\rho_S]$. Equations \eqref{e:AdjointMasterEquation} and \eqref{e:AdjointLindbladian} serve as the starting point for deriving the spin components' equations of motion in the Heisenberg picture. While this derivation is a standard procedure for closed quantum systems, the non-unitary nature of the dynamics generated by the Lindbladian introduces additional technical complications here. To illustrate this, we write $A(t)$ as $A(t)=\exp(t\Ldag)A$, in which case the adjoint master equation becomes
\begin{equation}\label{e:AdjointMasterEquation2}
	\partial_t A(t)=\Ldag A(t)=e^{t\Ldag}\Ldag A,
\end{equation}
where $\Ldag A$ is evaluated at $t=0$. While determining $\Ldag A$ is straightforward, the subsequent application of $\exp(t\Ldag)$ will not simply transform all the operators appearing in $\Ldag A$ into the Heisenberg picture. This is a consequence of the Lindbladian not satisfying the Leibniz rule, i.e.\ in general
\begin{equation}
	\Ldag(AB)\neq\Ldag(A)B+A\Ldag(B),
\end{equation}
and so
\begin{equation}
	e^{t\Ldag}(AB)\neq (e^{t\Ldag}A)(e^{t\Ldag}B).
\end{equation}
This complicates the task of extracting the desired equations of motion from the adjoint master equation. In what follows we show that, in the large-$S$ limit, the Lindbladian may be approximated to restore the Leibniz rule, without neglecting the dissipative effects of the bath coupling. Within this approximation the equation of motions can then be obtained easily.

Our approximation of the adjoint master equation in the $S\rightarrow \infty$ limit is based on the scaling behavior in $S$ of the various terms appearing in \eqref{e:AdjointLindbladian}. Within each term, every spin component contributes a factor of $S$ to the overall scaling, while every commutator contributes $S^{-1}$. This is a consequence of the fundamental commutator 
\begin{equation}\label{e:su2CR}
	[S_\mu,S_\nu]=i\epsilon_{\mu\nu\delta}S_\delta
\end{equation}
scaling linearly, rather than quadratically, in $S$. The general result for the commutator of arbitrary products of spin  components then follows from using the Leibniz rule to reduce the calculation to the repeated application of \eqref{e:su2CR}. Since all the operators of interest are derived from these components, we can use this simple rule to identify the dominant terms in \eqref{e:AdjointLindbladian}. After accounting for the $\mathcal{O}(S^{-1})$ scaling of the $\kappa$-coefficients, we find that the final three terms on the right of \eqref{e:AdjointLindbladian} are subdominant relative to the first two. Dropping these subdominant terms yields the approximation
\begin{equation}\label{e:ApproxAdjointLindbladian}
\Ldag \bullet\approx i[\mathcal{H},\bullet]+i\myIm(\kappa_{yx})\,\{S_y,[S_x,\bullet]\},
\end{equation}
for which it holds that
\begin{multline}\label{e:ApproxLeibniz}
	\Ldag(AB)\approx\Ldag(A)B+A\Ldag(B)\\
	+i\myIm(\kappa_{yx})\left([S_y,A][S_x,B]-[S_x,A][S_y,B]\right).
\end{multline}
The same scaling argument as before reveals that the term in the second line of \eqref{e:ApproxLeibniz} is subdominant in $S$. In the large-$S$ limit we may therefore take $\Ldag$ to satisfy the Leibniz rule, and so $e^{t\Ldag}$ will satisfy \mbox{$e^{t\Ldag}(AB)\approx (e^{t\Ldag}A)(e^{t\Ldag}B)$}. Combining this with \eqref{e:ApproxAdjointLindbladian} and inserting $\myIm(\kappa_{yx})=\gamma/(2S)$ turns the adjoint master equation in \eqref{e:AdjointMasterEquation2} into
\begin{equation}\label{e:AdjointMasterEquationFinal}
	\partial_t A(t)=i[\mathcal{H}_S(t),A(t)]+\frac{i\gamma}{2S}\,\left\{S_y(t),[S_x(t),A(t)]\right\},
\end{equation}
with all the operators now time-evolved in the Heisenberg picture. In Sec.~\ref{s:DissipativeDynamics} we will analyze the dynamics that follows from this approximated form of the master equation.

Note that when identifying the dominant terms in the original Lindbladian \eqref{e:AdjointLindbladian} we assumed that the temperature $T$ is intensive, and does not scale with system size. This resulted in the $\mathcal{O}(S^{-1})$ scaling of the $\kappa_{xx}$ coefficient, which led to the third term on the right of \eqref{e:AdjointLindbladian} falling way. Allowing for the temperature to scale extensively with $N$ will impact on both the dynamics of the observables and the equilibrium state thermodynamics. We return to this discussion in Sec.~\ref{ss:ExT}.

\section{Markovian dynamics of spin observables}
\label{s:DissipativeDynamics}
In this section we study the dynamics, generated by the approximated adjoint master equation in \eqref{e:AdjointMasterEquationFinal}, of the rescaled spin observables
\begin{equation}
	\mvec{s}=(x,y,z)\equiv(S_x,S_y,S_z)/S.
\end{equation}
We begin by deriving the equations of motion, which turn out to describe dynamics on two widely separated time scales. Using the two-timing method then allows us to analyze the equilibrium states and the energy dissipation rate at late times.

\subsection{Equations of motion \label{ss:com}}
Inserting the three spin components into  \eqref{e:AdjointMasterEquationFinal} and using the spin commutation relations $[S_\mu,S_\nu]=i\epsilon_{\mu\nu\delta}S_\delta$ leads to the equations of motion
\begin{subequations}
\begin{align}
	\dot{x}(t)&=y(t)\label{e:EOMx},\\
	\dot{y}(t)&=(\Lambda z(t)-1)x(t)-\gamma z(t) y(t),\label{e:EOMy}\\
	\dot{z}(t)&=-\Lambda x(t)y(t)+\gamma y^2(t).\label{e:EOMz}
\end{align}
\end{subequations}
In deriving these expressions we have used the fact that the commutators of the rescaled spin components are of order $\mathcal{O}(S^{-1})$, so that operator order does not matter. In the thermodynamic limit we can therefore treat $x$, $y$ and $z$ as commuting scalar variables which are constrained to the unit sphere as
\begin{equation}
	x^2+y^2+z^2=1, \label{e:UnitSphereConstraint}
\end{equation}
which is a reflection of the Casimir constraint $\mvec{S}^2=S(S+1)$. 

In the symmetric phase ($\Lambda\leq 1$) of the LMG model, the only stationary solution to \eqref{e:EOMx}--\eqref{e:EOMz} is  $(x,y,z)=(0,0,\pm 1)$, while in the symmetry-broken phase ($\Lambda>1$) there are two additional stationary points at $(x,y,z)=(\pm\sqrt{1-\Lambda^{-2}},0,\Lambda^{-1})$. To analyze how the dissipative dynamics steers the system towards these points we first introduce the rescaled system energy 
\begin{equation}
	h(t)=\mathcal{H}_S(t)/S=-\frac{\Lambda}{2}x(t)^2-z(t),\label{e:RescaledSystemEnergy}
\end{equation}
which, together with $x(t)$, obeys the set of coupled equations
\begin{subequations}
\begin{align}
 \dot{h}&=-\gamma \dot{x}^2,\label{e:EOMh}\\
 \ddot{x}&=-\frac{\Lambda^2}{2}x^3-(\Lambda h+1)x+\gamma\dot{x}\left(\frac{\Lambda}{2}x^2+h\right).\label{e:EOMx2}
\end{align}
\end{subequations}
From the expression for $\dot{h}$ we see that, as a result of the bath coupling, the system's energy decays on a time scale set by $\gamma$. For weak coupling, this will be much longer than the time scale associated with the dynamics of the individual spin components. This separation of time scales suggests the use of the two-timing method \cite{Strogatz}, which we implement in the next section. We focus for now on the evolution of $h$ and $x$, and return to the $y$ and $z$ components at the end.

\subsection{Two-timing method}
To capture the dependence of the observables on the two time scales that emerge at weak coupling, we introduce a fast time $\tau=t$ and a slow time $s=\gamma t$. These will be treated as independent variables. As an Ansatz for the solutions of \eqref{e:EOMh} and \eqref{e:EOMx2} we take
\begin{subequations}
\begin{align}
	x(t,\gamma)&=x_0(\tau,s)+\gamma\, x_1(\tau,s)+\mathcal{O}(\gamma^2),\\
	h(t,\gamma)&=h_0(\tau,s)+\gamma\, h_1(\tau,s)+\mathcal{O}(\gamma^2).
\end{align}
\end{subequations}
Inserting these forms into \eqref{e:EOMh} and \eqref{e:EOMx2} and collecting powers of $\gamma$ leads to
\begin{subequations}
\begin{align}
	\partial_\tau h_0&=0,\label{e:EOMh3}\\
	\partial^2_\tau x_0&=-\frac{\Lambda^2}{2}x_0^3-(1+\Lambda h_0)x_0,\label{e:EOMx3}\\
	\partial_s h_0&=-\partial_\tau h_1-(\partial_\tau x_0)^2.\label{e:EOMh4}
\end{align}
\end{subequations}
The first equation shows that $h_0(\tau,s)=h_0(s)$ is independent of $\tau$, and that the system's energy remains constant over the short time scale associated with $\tau$. This also implies that in \eqref{e:EOMx3} the $\tau$-dependence only enters through $x_0(\tau,s)$. The latter is the equation of motion for a particle moving in the effective quartic potential
\begin{equation}
	V(x)=\frac{\Lambda^2}{8}x^4+\frac{1}{2}(1+\Lambda h_0)x^2.\label{e:EffectivePotential}
\end{equation}
The solution for $x_0$ is given by \cite{NIST}
\begin{equation}
	x_0(\tau,s)=a\cn\left(\Omega \tau+u,k^2\right)\label{e:EllipticFunctionSolution}
\end{equation}
with
\begin{equation}
	\Omega=\sqrt{a^2\Lambda^2/2+h_0\Lambda+1},\qquad k^2=\frac{a^2\Lambda^2}{4\Omega^2},\label{e:DefineOmegaksquared}
\end{equation}
where $\cn$ denotes the Jacobi elliptic cosine function. Note that $a$ and $u$, which enter as integration constants in the solution for $x_0(\tau,s)$, are functions of $s$. To determine $a(s)$ we combine \eqref{e:EOMx}, \eqref{e:UnitSphereConstraint}, and \eqref{e:RescaledSystemEnergy} into an additional constraint on $x_0(\tau,s)$, which reads
\begin{equation}
	\left(\partial_\tau x_0\right)^2=1-(\Lambda x_0^2/2+h_0)^2-x_0^2. \label{e:constraintequation}
\end{equation}
This produces an equation for $a^2$ with solution
\begin{equation}
	a^2=\frac{2(-1-h_0\Lambda+\sqrt{\Lambda^2+2h_0\Lambda+1})}{\Lambda^2},
\end{equation}

which depends on $s$ through $h_0$. We may take $a$ to be positive, since a change in sign can be absorbed through an appropriate shift in $u$.

At this stage we have established the $\tau$-dependence of $x_0(\tau,s)$ and seen that the $s$ dependence enters via $h_0(s)$ and $u(s)$. We now turn to \eqref{e:EOMh4}, from which we extract the $s$-dependence of $h_0(s)$ in order to characterize the energy dissipation. First we average both sides of \eqref{e:EOMh4} with respect to $\tau$ over the interval $[0,\tau_f]$, yielding
\begin{equation}
	\partial_s h_0=\frac{h_1(0,s)-h_1(\tau_f,s)}{\tau_f}-\frac{1}{\tau_f}\int_0^{\tau_f}d\tau\,(\partial_\tau x_0)^2.\label{e:AveragedEOMh4}
\end{equation}
A crucial element in the implementation of the two-timing method is the elimination of secular terms which grow in $\tau$ without bound. Here this amounts to requiring that $h_1(\tau_f,s)/\tau_f\rightarrow 0$ as $\tau_f\rightarrow\infty$. Taking this limit in \eqref{e:AveragedEOMh4} eliminates the first term on the right-hand side and turns the second term into the long-time average of $(\partial_\tau x_0)^2$. The latter can be replaced by the average over a single period of the periodic solution for $x_0(\tau,s)$ in \eqref{e:EllipticFunctionSolution}. This produces
\begin{equation}
	\partial_s h_0=-\frac{1}{T(k)}\int_0^{T(k)}d\tau\,(\partial_\tau x_0)^2,\label{e:AverageOverPeriod}
\end{equation}
where the period of $x_0(\tau,s)$ is 
\begin{equation}
	T(k)=\begin{cases}\frac{4K(k^2)}{\Omega} & \text{for $k^2<1$}, \\ \infty & \text{for $k^2=1$}, \\ \frac{2K(k^{-2})}{\Omega k}& \text{for $k^2>1$},\end{cases}
\end{equation}
and $K(k^2)$ denotes the complete elliptic integral of the first kind \cite{NIST}. The averaging in \eqref{e:AverageOverPeriod} also eliminates any dependence on $u(s)$, and the result depends on $s$ only through $h_0(s)$. We write \eqref{e:AverageOverPeriod} compactly as
\begin{equation}
	\partial_s h_0(s)=-A(h_0(s))\label{e:EvolutionOfh0}
\end{equation}
where the result of the integral is
\begin{widetext}
\begin{equation}\label{e:Ah0}
	A(h_0)=\begin{cases}
	\frac{a^2\Omega^2}{3k^2}\left[(1-k^2)+(2k^2-1)\frac{E(k^2)}{K(k^2)}\right]&\text{for $k^2<1$},\\
	0&\text{for $k^2=1$},\\
    \frac{a^2\Omega^2}{3}\left[2(1-k^2)+(2k^2-1)\frac{E(k^{-2})}{K(k^{-2})}\right]&\text{for $k^2>1$}.
	\end{cases}
\end{equation}
\end{widetext}
Note that $k$ is itself a function of $h_0$ through \eqref{e:DefineOmegaksquared}. Equations \eqref{e:EvolutionOfh0} and \eqref{e:Ah0}, which are the main results of this subsection, describe the slow dissipation of the system's energy. Next we will reintroduce the operator character of $h_0(s)$ and study equilibration on the basis of Eq.~\eqref{e:EvolutionOfh0}.

\subsection{Approach to equilibrium}
\label{s:approach}
Since no other operators apart from $h_0(s)$ appear in \eqref{e:EvolutionOfh0}, $h_0(s)$ will remain diagonal in the original $h(0)$ basis. In this sense \eqref{e:EvolutionOfh0} describes the evolution of the eigenvalues of $h_0(s)$. Any such eigenvalue $\epsilon(s)$ satisfies 
\begin{equation}\label{e:EigenvalueEvolve}
	\partial_s \epsilon(s)=-A(\epsilon(s)),
\end{equation}
and the behavior of the eigenvalues at late times will reflect the system's equilibrium state. The possible initial eigenvalues $\epsilon(0)$ are determined by the range of the spectrum of $h(0)=-\Lambda x^2/2-z$. In the large-$S$ limit this range can be obtained by treating the rescaled spin components as commuting variables and finding the extremal values of $h(0)$ subject to the unit sphere constraint \eqref{e:UnitSphereConstraint}. It is found that $\epsilon(0)\in[\epsilon_g,1]$ where
\begin{equation}\label{e:spectralrange}
	\epsilon_g=\begin{cases} -1 & \text{for $\Lambda\leq 1$},\\ -\frac{1}{2}[\Lambda^{-1}+\Lambda] & \text{for $\Lambda>1$}. \end{cases}
\end{equation}
The same result follows from the standard variational approach based on spin coherent states \cite{DusuelVidal05}.

\begin{figure}[t]
	\centering
	\includegraphics[width=0.48\linewidth]{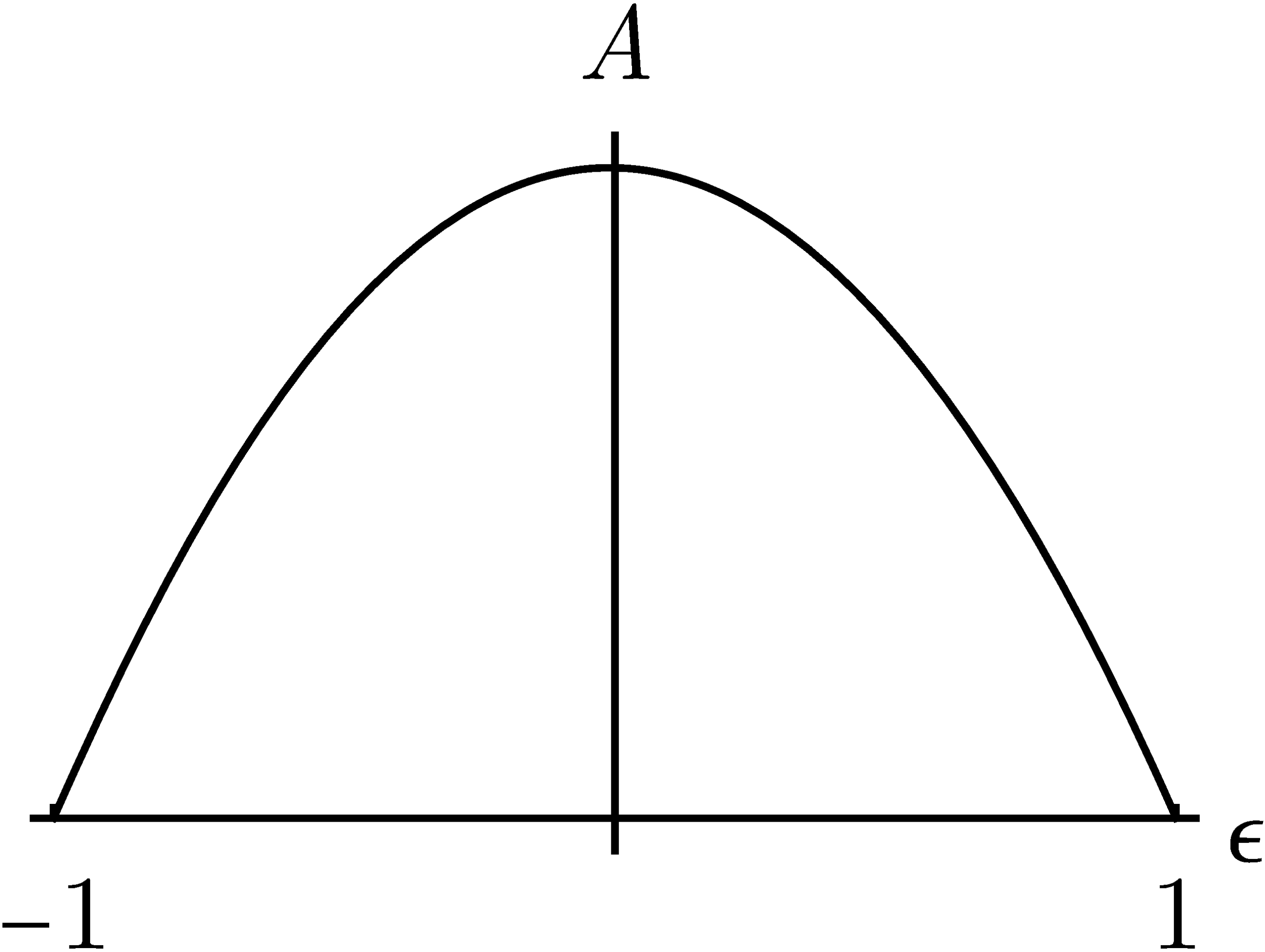}
	\hfill
	\includegraphics[width=0.48\linewidth]{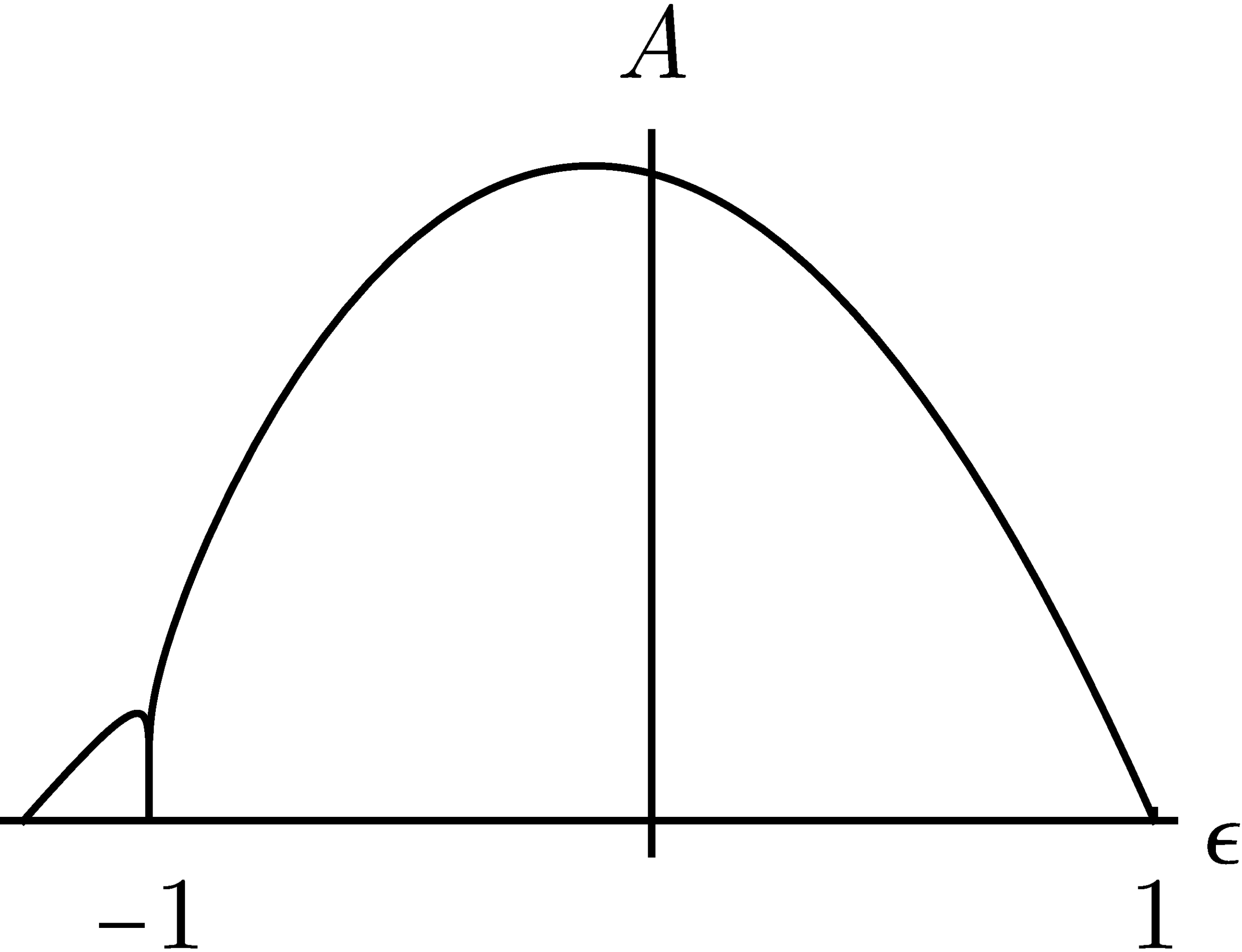}
	\caption[]{Plots of $A$ as a function of $\epsilon$ over the spectral range $[\epsilon_g,1]$ determined by \eqref{e:spectralrange}. According to Eq.~\eqref{e:EigenvalueEvolve}, a positive value of $A$ implies that $\epsilon$ evolves towards smaller values, while $A=0$ corresponds to a fixed point of the dynamics. Left: For $\Lambda=1/2$, $A$ is positive in the entire spectral range of $h_0$, except at the boundary points $\epsilon=\pm1$, implying that $\epsilon=-1$ is an attractive fixed point of the dynamics. Plots for other values of $\Lambda\leq1$ in the symmetric phase look qualitatively similar and lead to the same conclusions. Right: For $\Lambda=2$, $A$ is positive in the entire spectral range of $h_0$, except at the boundary points $\epsilon=-\epsilon_g$, $\epsilon=1$, and the interior point $\epsilon=-1$. As argued in the main text, the presence of this interior fixed point is an artifact of the large-$S$ limit. Plots for other values of $\Lambda>1$ in the symmetry-broken phase look qualitatively similar and lead to the same conclusions.}
	\label{fig:energysymbol2d}
\end{figure}

Figure \ref{fig:energysymbol2d} shows $A(\epsilon)$, which governs the evolution of $\epsilon(s)$, as a function of $\epsilon\in[\epsilon_g,1]$ for values of $\Lambda$ corresponding to the two phases. We see that $-A(\epsilon)$ is negative everywhere, except at the boundaries of the spectral range, and in the symmetry-broken phase also at $\epsilon=-1$. In the symmetric phase the eigenvalues starting in $[-1,1)$ therefore evolve towards $\epsilon_g=-1$, which is a stable fixed point of the evolution. In the symmetry-broken phase eigenvalues in $[-1,1)$ converge to $\epsilon=-1$, while those in $[\epsilon_g,-1)$ tend to $\epsilon_g$. The appearance of the fixed point at $\epsilon=-1$ for $\Lambda>1$ is an artifact of the large-$S$ limit and the expansion in $\gamma$. It corresponds to the scenario where the dissipation lowers $h_0(s)$ to precisely the value where $x_0(\tau,s)$, evolving according to \eqref{e:EOMx3}, gets ``stuck'' at the local maximum of the double-well potential $V(x)$ in \eqref{e:EffectivePotential}. While this is inevitable when considering only the dynamics of $h_0$ and $x_0$, it clearly requires very specific initial conditions to occur in the original equations of motion of $h$ and $x$ in \eqref{e:EOMh} and \eqref{e:EOMx2}. The $\epsilon=-1$ fixed point is therefore not physically significant, and will not halt the system's evolution at finite $S$, nor when the coupling $\gamma$ is treated non-perturbatively. The relevant fixed point in the symmetry-broken phase is therefore $\epsilon_g$, which  serves as an attractor for all initial eigenvalues in $[\epsilon_g,1)$. Returning to the operator level, we conclude that $\lim_{s\rightarrow\infty} h_0(s)=\epsilon_0\hat{I}$. Combining this with \eqref{e:RescaledSystemEnergy} and \eqref{e:constraintequation} implies the equilibrium expectation values
\begin{equation}
	\ev{x_0^2}_\text{eq}={\rm max}\{0,1-\Lambda^{-2}\},\quad \ev{z_0}_\text{eq}={\rm min}\{\Lambda^{-1},1\},
\end{equation}
and $\ev{y_0}_\text{eq}=0$. These coincide with the stable equilibria of \eqref{e:EOMx}--\eqref{e:EOMz} identified earlier.

To determine the dissipation rate close to equilibrium we consider the expectation value of $h_0(s)$ with respect to an arbitrary initial state.  This expectation value evolves according to 
\begin{equation}
	\partial_s \ev{h_0}(s)=\ev{-A(h_0(s))}.\label{e:EvolutionOfh0EV}
\end{equation}
With all the eigenvalues of $h_0(s)$ tending towards $\epsilon_0$, we can at late times approximate $\ev{-A(h_0(s))}$ by $-A(\ev{h_0}(s))$. The behavior of $A(\epsilon)$ around $\epsilon=\epsilon_g$ is
\begin{equation}
	A(\epsilon)=\begin{cases} (\epsilon-\epsilon_g)+\mathcal{O}((\epsilon-\epsilon_g)^2) & \text{for $\Lambda<1$},\\ \frac{4}{3}(\epsilon-\epsilon_g)+\mathcal{O}((\epsilon-\epsilon_g)^{3/2}) & \text{for $\Lambda=1$},\\ \frac{1}{\Lambda}(\epsilon-\epsilon_g)+\mathcal{O}((\epsilon-\epsilon_g)^2) & \text{for $\Lambda>1$}.\end{cases}
\end{equation}
Close to equilibrium we therefore have
\begin{equation}
	\partial_t \ev{h_0}(t)=-\omega_h(\ev{h_0}(t)-\epsilon_g),\label{e:EvolutionOfh0EV2}
\end{equation}
where 
\begin{equation}
	\omega_h=\gamma \begin{cases} 1 & \text{for $\Lambda<1$},\\ \frac{4}{3} & \text{for $\Lambda=1$},\\ \frac{1}{\Lambda} & \text{for $\Lambda>1$},\end{cases}
\end{equation}
is the phase-dependent dissipation rate. Note the switch from rescaled time $s=\gamma t$ back to $t$. Solving \eqref{e:EvolutionOfh0EV2} for $\ev{h_0}(t)$ yields
\begin{equation}
	\ev{h_0}(t)=\epsilon_g+Ce^{-\omega_h t}
\end{equation}
with $C$ a constant.
	
\section{Extensive temperatures}
\label{ss:ExT}
We have shown that at any fixed finite temperature the equilibrium expectation values of intensive quantities match their ground state values in the large-$S$ limit. This does not imply thermalization to the ground state, but rather that these quantities cannot resolve the difference between the ground state and the true mixed equilibrium state. To observe equilibration towards states for which, even in the large-$S$ limit, these expectation values deviate from those of the ground state requires, as we will see, the temperature $T$ to be considered an extensive quantity. This is equivalent to taking the large-$S$ limit at a fixed rescaled temperature $\widetilde{T}=T/S$. In the Lindbladian \eqref{e:Lindbladian} this amounts to replacing $\kappa_{xx}$ by $2\gamma\widetilde{T}$ \footnote{We consider $\nu_1$, and hence $\kappa_{xy}$ and $\kappa_{yx}$, as asymptotically independent of $S$. Since $\nu_1$ in \eqref{eq:nu1def} grows proportional to $T/\omega_c$, this essentially amounts to rescaling the cut-off frequency $\omega_c\to S\omega_c$ of the bath spectral density \eqref{e:LorentzDrude}.}, yielding
\begin{multline}\label{eq:RspinCom}
\mathcal{L}\bullet= i[\bullet,\mathcal{H}_S]-\frac{i\gamma}{2S}[S_x,\{S_y,\bullet\}]-\gamma\widetilde{T}[S_x,[S_x,\bullet]]\\
- \frac{\gamma(\nu_1^2+1/4)}{2\widetilde{T}S^2}[S_y,[S_y,\bullet]] + \frac{2\gamma\nu_1}{S} [S_y,[S_x,\bullet]].
\end{multline}
Compared to Eq.~\eqref{e:AdjointMasterEquationFinal}, the last three terms on the right-hand side make the master equation \eqref{eq:RspinCom} more complicated to deal with, and we were not able to analyse the long-time dynamics along the lines of Sec.~\ref{s:DissipativeDynamics}. As an alternative approach, we show in this section analytically that a Gibbs thermal state is a stationary state of this master equation in the large-$S$ limit, and provide numerical evidence that no other stationary states exist. Trace preservation of the master equation then implies that any initial state evolves towards the unique stationary state, which establishes thermalization.

We consider the Gibbs thermal state 
\begin{equation}\label{e:Gibbs}
\varrho_{\widetilde{\beta}}=\frac{e^{-\beta \Hh_S}}{Z} = \frac{e^{-\widetilde{\beta} h}}{Z}
\end{equation}
with $h=\mathcal{H}_S/S$, where $\widetilde{\beta}=S\beta=1/\widetilde{T}$ is the rescaled inverse temperature and $Z=\Tr[\exp(-\beta \Hh_S)]$ is the partition function. To establish the stationarity of the master equation \eqref{e:MasterGeneral} in the large-$S$ limit we compare the scaling in $S$ of $\dot{\varrho}_{\widetilde{\beta}}=\Ll\varrho_{\widetilde{\beta}}$ with that of $\varrho_{\widetilde{\beta}}$ itself. Here the linearity of $\Ll$ allows us to replace $\varrho_{\widetilde{\beta}}$ by $e^{-\widetilde{\beta} h}$ for the sake of this comparison. We will show that, while $e^{-\widetilde{\beta} h}=\Oo(S^{0})$, a delicate cancellation of leading order terms results in $\Ll e^{-\widetilde{\beta} h}$ being of order $\Oo(S^{-1})$. In this sense the relative rate of change of $\varrho_{\widetilde{\beta}}$ becomes negligible in the large-$S$ limit.

Inserting $e^{-\widetilde{\beta} h}$ into \eqref{eq:RspinCom}, the first commutator on the right-hand side of \eqref{eq:RspinCom} is zero. The arguments used to identify the dominant terms in \eqref{e:AdjointLindbladian} for the adjoint Lindbladian, which eventually led to \eqref{e:ApproxAdjointLindbladian}, can likewise be employed to determine the scaling behaviour of the remaining terms in \eqref{eq:RspinCom}. The final two terms are of order $\Oo(S^{-2})$ and $\Oo(S^{-1})$, respectively, while the second and third terms are individually of order $\Oo(S^{0})$.  As we show next, when $\mathcal{L}$ acts on $e^{-\widetilde{\beta} h}$ there are cancellations between the latter two terms which result in their combined contribution being of order $\Oo(S^{-2})$. For each of these terms, we first combine the prefactor with the innermost (anti)commutator, which can be expanded in a power series in $1/S$ using the BCH formula. For the second term in \eqref{eq:RspinCom} this yields
\begin{align}\label{e:exp1}
	-\frac{i\gamma}{2S}\{S_y,e^{-\widetilde{\beta} h}\}&=-\frac{i\gamma}{2}\left[y+e^{-\widetilde{\beta} h}ye^{\widetilde{\beta} h}\right]e^{-\widetilde{\beta} h}\\
	&=-i\gamma\left[y-\frac{\widetilde{\beta}}{2}[h,y]+\Oo(S^{-2})\right]e^{-\widetilde{\beta} h},\nonumber
\end{align}
where the omitted terms involve further nested commutators with $h$, each of which contributes a factor of $S^{-1}$ to the term's overall scaling. Applying the same procedure to the third term in \eqref{eq:RspinCom} produces 
\begin{align}\label{e:exp2}
	-\gamma\widetilde{T}[S_x,e^{-\widetilde{\beta} h}]&=-\gamma\widetilde{T}S\left[x-e^{-\widetilde{\beta} h}xe^{\widetilde{\beta} h}\right]e^{-\widetilde{\beta} h}\\
	&=i\gamma\left[y-\frac{\widetilde{\beta}}{2}[h,y]+\Oo(S^{-2})\right]e^{-\widetilde{\beta} h}\nonumber
\end{align}
which, up to order $\Oo(S^{-1})$, differs from the previous result only by a sign. Inserting \eqref{e:exp1} and \eqref{e:exp2} into the sum of the second and third term of \eqref{eq:RspinCom} then yields a combined contribution of the form $[S_x,\Oo(S^{-2})e^{-\widetilde{\beta} h}]$, which is itself of order $\Oo(S^{-2})$. Taken together, these arguments establish the $\Oo(S^{-1})$ scaling of $\Ll e^{-\widetilde{\beta} h}$ and therefore also the stationarity of $\varrho_{\widetilde{\beta}}$ in the large-$S$ limit.

\begin{figure}[t]
	\centering
	\includegraphics[width=\linewidth]{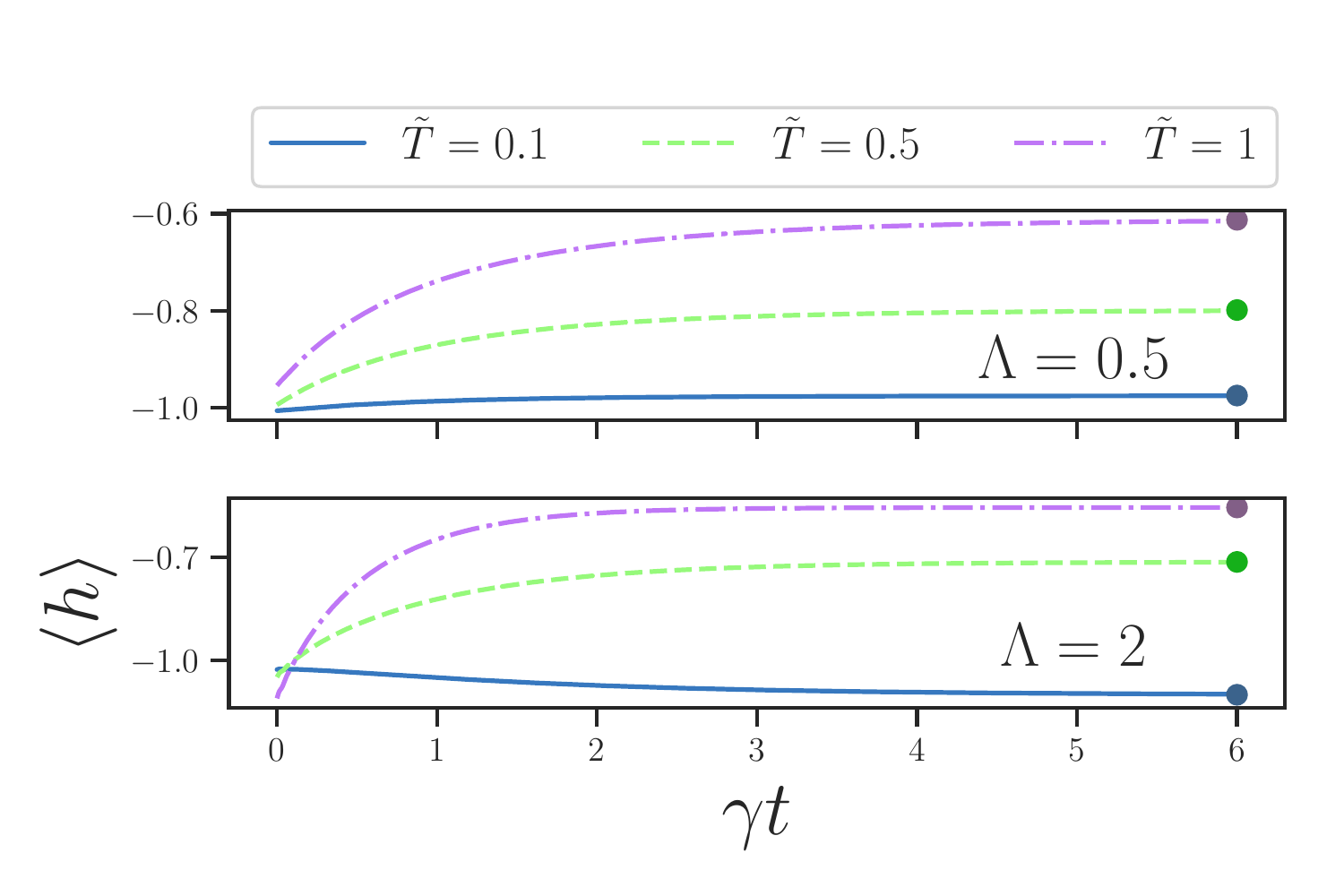}
	\caption[]{Plots of $\ex{h}$ as a function of rescaled time $\gamma t$ for initial states with different initial energies, and for different extensive temperatures $\widetilde{T}$. The top plot is for $\Lambda = 1/2$, corresponding to the symmetric phase, the bottom plot for $\Lambda= 2$ in the symmetry-broken phase. The dots mark the thermal equilibrium values of the energy densities at the respective temperatures, calculated using the Gibbs state \eqref{e:Gibbs}.}
	\label{fig:energy}
\end{figure}

The uniqueness of the stationary state $\varrho_{\tilde\beta}$ is encoded in the spectrum of the Lindbladian \eqref{eq:RspinCom}. Denoting the eigenvalues of $\mathcal{L}$ by $\lambda_i$ and the corresponding eigenstates by $\varrho_i$, it follows that $e^{t\Ll} \varrho_i = e^{t \lambda_i} \varrho_i$. Stationary states correspond to $\lambda_i = 0$, while states with $\myRe\lambda_i<0$ decay in time. We order the eigenstates by their real part such that $\myRe \lambda_i \ge \myRe\lambda_{i+1}$ for all $i$. The stationary state associated with $\lambda_0=0$ is unique if $\myRe\lambda_1 < 0$. Using the \emph{Quantum Toolbox in Python} \cite{Johansson_etall} we calculated the gap $\lambda_0-\lambda_1$ of the Lindbladian for system sizes up to $S = 80$. We find a nonzero gap that is essentially independent of $S$, supporting the conclusion that the gap persists in the large-$S$ limit. Accepting this conclusion as a fact, it follows that the Gibbs thermal state \eqref{e:Gibbs} is the unique stationary state, implying thermalization under the dynamics of the master equation \eqref{e:MasterGeneral} with Lindbladian \eqref{eq:RspinCom}. In Fig.~\ref{fig:energy} this conclusion is corroborated by numerical calculations of the time evolution of the energy expectation value $\ex{h}$, showing convergence to the predicted thermal equilibrium value independently of the initial state.

\section{Conclusions}

We derived a Lindblad master equation that approximates the dynamics of an LMG model weakly coupled to a bosonic bath of temperature $T$. The derivation mostly follows the standard procedure of performing Born and Markov approximations \cite{BreuerPetruccione}, but avoids the secular approximation, which is usually the final step of the derivation. Avoiding this step results in a time evolution equation that is not completely positive. We restore complete positivity by a minimal modification of the jump rates, see Eq.~\eqref{e:kappa_def} and below, to obtain a master equation that satisfies the requirements of a quantum dynamical semi-group.

In the absence of a bath, the LMG model is integrable and does not equilibrate in general. On the basis of our Lindblad master equation we analyzed whether the presence of a bath indeed enforces, as on would expect, thermalization, i.e., equilibration to a Gibbs state of temperature $T$ independently of the system's initial state.

Our first main result establishes that, for an LMG model in the large-system (or large-$S$) limit, thermalization occurs in the sense that expectation values of the system's Hamiltonian density $\mathcal{H}_S/S$ as well as expectation values of all three spin component densities $S_x/S$, $S_y/S$, and $S_z/S$ converge in the long-time limit to their thermal equilibrium values. A peculiar property of those thermal equilibrium values is that they agree, for all values of the bath temperature $T$, with the ground state expectation values of the LMG model. This peculiarity of the model was previously discussed in \cite{WebsterKastner18} in the equilibrium context, and is confirmed in a dynamical context by the results of the present paper.

The unconventional thermal behavior of the LMG model is a consequence of the small Hilbert space dimension of the model. While in conventional spin models the Hilbert space dimension grows exponentially with the system size $N$, it grows only linearly with $N$ (or $S$) in the case of the LMG model. Due to the small number of available states, entropic effects are mostly irrelevant and the effect of temperature, which is conjugate to the entropy, is suppressed.

A strategy to force the system into highly excited thermal equilibrium states even in the absence of entropic effects consists in promoting temperature to become an extensive quantity. This is achieved formally by substituting $T\to\tilde{T}S$ in the master equation and considering $\tilde{T}$ to be constant while taking the large-$S$ limit. This procedure renders terms in the master equation that were previously subleading in $1/S$ into leading ones. Analyzing the resulting equation, we found in Sec.~\ref{ss:ExT} that a Gibbs thermal state with rescaled temperature $\tilde{T}$ is stationary. Establishing thermalization requires to show that this Gibbs state is the only stationary state. We were not able to prove uniqueness of the stationary state analytically, but provided numerical evidence in favor of this claim.

Rescaling temperature with $S$ may seem as an artificial and unphysical procedure, but this is not the case. The large-system limit, or large-$S$ limit, is an idealization, and one is ultimately interested in large but finite systems. Thermalization in rescaled temperature $\tilde{T}$ then states that a large system of size $S$ and (unscaled) temperature $T$ will have a thermal equilibrium state very similar to that of a system of size $2S$ at temperature $2T$, etc., and the limit of such a sequence is given by a Gibbs state with rescaled temperature.

All results of this paper have been proved for a master equation that was derived for a bosonic bath with a Lorentz-Drude spectral density \eqref{e:LorentzDrude} and a system--bath coupling \eqref{eq:coup}. It is unclear at present how the specific properties of an environment might look like in an experimental realization of the LMG model in a bath, but one would hope that, as in other known cases, details of the bath are not overly important and our results carry over to a larger class of environments, including experimentally relevant ones. This would then open up an avenue for imposing thermalization on the otherwise non-thermalizing LMG model and for probing its unconventional thermodynamic properties.

\acknowledgments
M.\,K.\ gratefully acknowledges helpful discussions with I.~de Vega, hospitality at the Kavli Institute for Theoretical Physics at Santa Barbara during the QSIM19 program, and financial support by the South African National Research Foundation through the Incentive Funding Programme and the Competitive Funding for Rated Researchers. This research was supported in part by the National Science Foundation under Grant No.~NSF PHY-1748958.

\appendix

\section{Born-Markov master equation}
\label{A:BMM}
For the derivation of the master equation we consider the von Neumann equation in the interaction picture,
\begin{equation}\label{Neumann}
\dot{\varrho}^{(I)}_{SB}(t) = i \left[ \varrho^{(I)}_{SB}(t), \mathcal{V}(t)\right],
\end{equation}
where $\mathcal{V}(t) = U(t)\mathcal{V} U^{\dag}(t)$ with $U(t) = \text{exp}(i\mathcal{H}_0 t)$. Assuming system and bath to be uncorrelated at $t=0$, system--bath interactions to be weak, and the bath to evolve much faster than the system, we employ a sequence of standard approximations \cite{BreuerPetruccione} to derive the Born-Markov master equation 
\begin{equation}\label{eq:DEBA3}
\dot{\varrho}_S^{(I)}(t) = \frac{\gamma}{S} \int_0^\infty d\tau \ex{\mathcal{B}(\tau)\mathcal{B}} [ S_{x}(t-\tau)\varrho_S^{(I)}(t), S_{x}(t)] + \text{h.c.},
\end{equation}
where $\text{h.c.}$ denotes the hermitian conjugate. Transforming \eqref{eq:DEBA3} back into the Schr{\"o}dinger picture one obtains
\begin{equation}\label{eq:diff1}
\dot{\varrho}_S(t) = i[\varrho_S(t), \mathcal{H}_S + \mathcal{H}_c] + ([S_{\mathcal{B}} \varrho_S(t), S_{x}] + \text{h.c.}),
\end{equation}
where we have defined the counter term
\begin{equation}
\mathcal{H}_c =  \frac{\gamma}{S} S_x^2 \sum_{q} \frac{\gamma_q^2}{\omega_q}
\end{equation}
and the coupling operator
\begin{equation}\label{eq:SBb}
S_\mathcal{B} \equiv \frac{\gamma}{S} \int_0^\infty d\tau \ex{\mathcal{B}(\tau)\mathcal{B}}_0 S_{x}(-\tau)
\end{equation}
with
\begin{equation}\label{e:bathcorr}
\ex{\mathcal{B}(\tau) \mathcal{B}}_0 \equiv \Tr\left[\varrho_B(0) \mathcal{B}(\tau) \mathcal{B}\right].
\end{equation}
It is common to simplify \eqref{eq:diff1} further by performing a secular or rotating-wave approximation, which amounts to averaging over rapidly oscillating terms in the master equation. This approximation guarantees that the resulting master equation is a quantum dynamical semi-group \cite{BreuerPetruccione}. However, the approximation also imposes, possibly artificially, a Gibbs state to be a stationary state of the dynamics \cite{SpohnLebowitz07,BreuerPetruccione}. Further this approximation fails close to a phase transition due to the divergence in time scales \cite{Koplov_etal}. Here we continue without a secular or rotating-wave approximation, and we prove in Appendix \ref{ssec:pos} that, after some tweaking, our master equation nonetheless satisfies the desired properties of a quantum dynamical semi-group.

To simplify the master equation \eqref{eq:diff1} it is convenient to separate the bath correlation function \eqref{e:bathcorr} into real and imaginary parts. Assuming the bath to be in a Gibbs state with inverse temperature $\beta$, one can calculate the so-called decoherence kernel to be \cite{Schlosshauer}
\begin{equation}\label{eq:nueta1}
\nu(\tau) \equiv \myRe \ex{\mathcal{B}(\tau) \mathcal{B}}_0= \sum_{q} \gamma_q^{2} \coth\left(\frac{\beta \omega_q}{2}\right) \cos(\omega_q \tau)
\end{equation}
and the noise kernel to be
\begin{equation}\label{e:eta1}
\eta(\tau) \equiv -\myIm \ex{\mathcal{B}(\tau) \mathcal{B}}_0= \sum_{q} \gamma_q^{2} \sin(\omega_q \tau).
\end{equation}
Irreversible dynamics, and hence equilibration in an asymptotic sense, can occur only for an infinite bath with a continuous frequency spectrum \cite{BreuerPetruccione}. Formally we promote the set of discrete frequencies $\omega_q$ to a continuous variable $\omega$ and replace $\sum_q \gamma_q^2$ by $\int d\omega J(\omega)$ in \eqref{eq:nueta1} and \eqref{e:eta1}. We choose a spectral density $J$ of Ohmic type with a Drude cut-off \cite{BreuerPetruccione},
\begin{equation}\label{e:LorentzDrude}
J(\omega) = \frac{\omega}{\pi}  \frac{\omega_c^2}{\omega_c^2 + \omega^2},
\end{equation}
where the cut-off frequency $\omega_c$ should be chosen such that the dynamics becomes essentially independent of $\omega_c$. After substitution one finds
\begin{equation}\label{eq:eta1}
\eta(\tau) = \int_0^\infty d\omega \frac{1}{\pi} \omega \frac{\omega_c^2}{\omega_c^2 + \omega^2} \sin(\omega \tau)= \omega_c^2 e^{-\omega_c \tau}/2
\end{equation}
for the noise kernel and
\begin{equation}\label{eq:nu1}
\nu(\tau) = \frac{\omega_c^2}{2} e^{-\omega_c \tau}  \cot\left(\frac{\beta\omega_c}{2}\right) -\frac{\omega_c^2}{\pi} \sum_{n=1}^{\infty} \frac{|n| e^{-2\pi\tau |n|/\beta}}{(\beta\omega_c/(2 \pi))^2 - n^2}
\end{equation}
for the decoherence kernel \cite{BreuerPetruccione}, both of which decay exponentially with $\tau$. The corresponding decay rates are $\omega_c$ and $2\pi/\beta$, such that the bath correlation time is
\begin{equation}
\tau_B = \max\{ 1/\omega_c, \beta /(2 \pi)\}.
\end{equation}

To simplify the master equation \eqref{eq:DEBA3} we want to evaluate the integral \eqref{eq:SBb}. The integrand contains $S_x(-\tau)$, and while an exact solution of this time-evolved operator seems out of reach, the first few terms of the Baker-Campbell-Hausdorff formula  give the approximate solution
\begin{align*}
S_{x}(-\tau) &=  S_{x} - i[\mathcal{H}_S + \Oo(\gamma), S_{x}] \tau + \Oo(\tau^2)\\
&=  S_{x}-\tau  S_y + \Oo(\tau^2) + \Oo(\gamma),\label{eq:BCHSx}
\numberthis\end{align*}
valid for small bath correlation times $\tau_B$. Using the approximate solution \eqref{eq:BCHSx} together with \eqref{eq:eta1} and \eqref{eq:nu1} one can show that \eqref{eq:SBb} simplifies to \cite{BreuerPetruccione}
\begin{equation}
S_\mathcal{B} = \frac{\gamma}{S} \left[\left(T - \frac{i\omega_c}{2}\right) S_{x} - \left(\nu_1 - \frac{i}{2}\right)  S_y \right], \label{eq:Sbinted}
\end{equation}
where higher-order terms in $\tau$ and $\gamma$ have been neglected. Here we have defined
\begin{equation}\label{eq:nu1def}
\nu_1(q) \equiv \int_0^\infty d\tau \nu(\tau) \tau = \frac{T}{\omega_c} - \frac{\psi(\beta \omega_c/(2 \pi)+1)-\psi(1)}{\pi},
\end{equation}
where $\psi$ denotes the digamma function \cite{NIST}. 

Using \eqref{eq:Sbinted} we can now write the master equation \eqref{eq:diff1} in the form
\begin{equation}\label{e:Lindblad}
\dot{\varrho}_S(t) = (\Uu+ \Dd)\varrho_S(t)
\end{equation}
with a unitary part
\begin{equation}
\Uu \bullet = i[\bullet, \mathcal{H}_S + \mathcal{H}_\gamma]
\end{equation}
where
\begin{equation}
\mathcal{H}_\gamma \equiv \frac{1}{2 i} \left( S_x  S_\mathcal{B} -  S_\mathcal{B}^\dag  S_x\right) + \mathcal{H}_c 
 \approx \frac{\gamma}{4 S} \left\{ S_x,  S_y\right\} + \Oo(S^0),\label{e:Hgamma}
\end{equation}
and a dissipative part
\begin{align}
\Dd \bullet \equiv S_\mathcal{B} \bullet  S_{x} - \frac{1}{2} \{ S_{x} S_\mathcal{B},\bullet\} + \text{h.c.}\label{eq:dis}
\end{align} 
Inserting \eqref{eq:Sbinted} into \eqref{eq:dis} one finds that all terms containing $\omega_c$ cancel. This justifies to use in \eqref{eq:dis} the simpler coupling operator
\begin{equation}\label{eq:Sbprime}
S_\mathcal{B}' = \frac{\gamma}{S}  T S_x + \frac{\gamma}{S}\frac{i }{2}  S_y + \frac{\gamma}{S}  \nu_1 S_y,
\end{equation}
as it leads to an identical master equation.

\section{Lindbladian}
\label{ssec:pos}
In this section we make adjustments to the master equation \eqref{e:Lindblad} that impose the properties of a quantum dynamical semi-group onto the dynamics, which guarantees that a quantum mechanical density operator remains a density operator under time evolution for all times and initial states. Lindblad \cite{Lindblad76} has shown that this is guaranteed if $\Uu$ and $\Dd$ in \eqref{e:Lindblad} can be written as
\begin{equation}
\Uu\bullet = i[\bullet,\mathcal{H}], \qquad \Dd\bullet = \sum_{k,l} \kappa_{kl} \left(L_k \bullet L_l^{\dag} - \tfrac{1}{2} \{L_l^{\dag}L_k,\bullet\} \right) \label{eq:GKSL}
\end{equation}
and satisfy the following conditions:
\begin{enumerate}
\renewcommand{\labelenumi}{(\roman{enumi})}
\item $\mathcal{H}$ is hermitian. \label{a}
\item $L_k$ are traceless and pairwise orthonormal operators with respect to the Hilbert-Schmidt inner product, $\Tr\{L_k^{\dag} L_l\} = \delta_{kl}$. \label{b}
\item The matrix $\kappa$ with elements $\kappa_{kl}$ is hermitian and positive semi-definite. \label{c}
\end{enumerate}
For our master equation, (i) is clearly satisfied by \eqref{e:Hspin} and \eqref{e:Hgamma}. To satisfy (ii) we choose $L_1=S_x/S$ and $L_2=S_y/S$. 
To assess condition (iii) we write the dissipator \eqref{eq:dis} with coupling operator \eqref{eq:Sbprime} as
\begin{equation}\label{e:DissXY}
\Dd\bullet = \sum_{k,l \in \{x,y\}} \kappa_{kl} \left( S_k \bullet  S_l - \tfrac{1}{2} \{ S_l S_k,\bullet\} \right)
\end{equation}
with
\begin{equation}\label{e:kappa_def}
\kappa \equiv \begin{pmatrix}
\kappa_{xx} & \kappa_{xy}\\
\kappa_{yx} & \kappa_{yy}
\end{pmatrix} = \frac{\gamma}{2S}\begin{pmatrix}
4 T & (2\nu_1-i)\\
(2\nu_1+i) & 0
\end{pmatrix}.
\end{equation}
The matrix $\kappa$ in \eqref{e:kappa_def} is Hermitian, but not positive semi-definite, as is evident from
\begin{equation}
\det \kappa =\kappa_{xx}\kappa_{yy} - |\kappa_{xy}|^2 <0.\label{eq:detK}
\end{equation}
To restore positivity, we follow section 3.6.2.1 of \cite{BreuerPetruccione} and determine the minimal modification of $\kappa_{yy}$ that leads to $\det{\kappa} = 0$. From \eqref{eq:detK} we read off that this is achieved by setting $\kappa_{yy} = |\kappa_{xy}|^2/\kappa_{xx}$. Modifying $\kappa_{yy}$ ``by hand'', while not strictly justified, is expected to be reasonable at least when $|\kappa_{xy}|^2/\kappa_{xx}=(\nu_1^2+1/4)/T$ is small. Writing $\kappa$ in its spectral decomposition we note that the zero-eigenvalue term comes with a prefactor of zero. In that basis it is therefore sufficient to consider, instead of \eqref{e:DissXY}, the dissipator
\begin{equation}
\Dd \bullet = L \bullet L^\dag - \frac{1}{2} \{L^\dag L, \bullet\}
\end{equation}
with a single jump operator
\begin{equation}
L  = \sqrt{\kappa_{xx}} \left( S_{x} +\frac{\kappa_{yx}}{\kappa_{xx}}  S_y\right), \label{eq:firstjump}
\end{equation}
which can be left unnormalized.
After some tedious manipulations that lead to cancellations between the unitary and dissipative terms of the master equation, we obtain the final Born-Markov master equation \eqref{e:MasterGeneral} with Lindbladian \eqref{e:Lindbladian}, where $\kappa_{xx}$, $\kappa_{xy}$, and $\kappa_{yx}$ are defined as in \eqref{e:kappa_def}.
\bibliography{MK}

\end{document}